# RESONANT PLANETARY DYNAMICS: PERIODIC ORBITS AND LONG-TERM STABILITY.

George Voyatzis, Kyriaki I. Antoniadou and John D. Hadjidemetriou

Section of Astrophysics, Astronomy and Mechanics
Department of Physics, Aristotle University of Thessaloniki
54124 Thessaloniki, Greece
e-mail: voyatzis@auth.gr,  web page: http://users.auth.gr/voyatzis

**Abstract.** *Many exo-solar systems discovered in the last decade consist of planets orbiting in resonant configurations and consequently, their evolution should show long-term stability. However, due to the mutual planetary interactions a multi-planet system shows complicated dynamics with mostly chaotic trajectories. We can determine possible stable configurations by computing resonant periodic trajectories of the general planar three body problem, which can be used for modeling a two-planet system. In this work, we review our model for both the planar and the spatial case. We present families of symmetric periodic trajectories in various resonances and study their linear horizontal and vertical stability. We show that around stable periodic orbits there exist regimes in phase space where regular evolution takes place. Unstable periodic orbits are associated with the existence of chaos and planetary destabilization.*

## 1 INTRODUCTION

The three body problem (TBP) is a famous problem since it is very simple, yet with complex dynamics. Also, it is a very important problem for astronomy, since it can be used for the study of many celestial systems. In this work, we study the dynamics of planetary systems consisting of two planets, which can be modeled by the TBP. Our study is focused on the determination of periodic orbits and their stability. Generally, periodic orbits are associated with dynamical resonances, which, in the case of planetary systems, also correspond to planetary orbits with commensurate physical periods, i.e. $pT_1 + qT_2 = 0$, where $p,q$ are prime integers[1]. Such resonances are very common in the recently discovered extrasolar systems. The planetary migration, which takes place in the early stages of planetary formation, seems to be affected strongly by resonances[2,3] and the planets end in a resonant stable configuration, i.e nearby a stable periodic orbit.

As far as periodic orbits of the TBP are concerned, the majority of literature is devoted to the simplest case, which is the planar circular restricted TBP[4]. The planar elliptic and the circular three dimensional models have been also studied and applied to asteroid dynamics[5,6]. For planetary systems, the mutual gravitational interactions between the planets are generally important and therefore the suitable model is the general TBP. For the planar case, many resonances have been studied and many families of periodic orbits have been determined[7,8]. It has been shown that when dissipative forces act to the planetary system, the system migrates along such families of periodic orbits[3].

The three dimensional general TBP had been also studied many years ago and the conditions for the existence and continuation of periodic orbits had been given but only few families were computed which, generally, were not associated with planetary dynamics[9]. Recently, the planetary 2/1 resonance in space has been studied in detail and many stable branches of families of periodic orbits have been given. The study showed that there exist stable planetary configurations where the planets show large values of mutual inclinations[10].

In this work, we present how the TBP is configured to model the dynamics of a spatial two-planet system. We discuss the main concepts about periodic orbits and present results from our study of the 3/1, 5/2, 3/2 and 4/1 resonances.

## 2 THE ROTATING FRAME MODEL AND PERIODIC CONDITIONS

### 2.1 The TBP in a rotating frame

Let two planets $P_1$ and $P_2$ revolve around a star $P_0$. All bodies are assumed as point masses, $m_1$, $m_2$ and $m_0$, respectively, with stellar mass $m_0 \gg m_i$, $i$=1,2. The three bodies move in space O$XYZ$ (inertial frame) under their mutual gravitational forces and the equations of motion are

$$\ddot{\mathbf{R}}_i = -G\sum_{j=0}^{2}\frac{m_j}{r_{ij}^{3/2}}(\mathbf{R}_i - \mathbf{R}_j), \qquad i \neq j, \quad i = 0,1,2. \tag{1}$$

where $\mathbf{R}_i$ are the position vectors of the bodies, $r_{ij} = |\mathbf{R}_i - \mathbf{R}_j|$ and G is the gravitational constant, which is set equal to unity ($G=1$) for normalization reasons. By placing the origin O at the center of mass of the system we get $\sum_{i=1}^{3} m_i \mathbf{R}_i = 0$ and $\sum_{i=1}^{3} m_i \dot{\mathbf{R}}_i = 0$ and, therefore, we can determine the position and the velocity of the star $P_0$ via those of the two planets. Thus, the system in the inertial frame OXYZ is of 6 degrees of freedom and possesses the integrals of energy $E$ and angular momentum $\mathbf{L} = L_x \mathbf{i} + L_y \mathbf{j} + L_z \mathbf{k}$.

We can reduce the number of degrees of freedom by introducing a suitable frame of reference[9]. Firstly, we assume that $OZ \parallel \mathbf{L}$ and consider the inertial frame $GX'Y'Z'$ after a parallel translation $\mathsf{T}(\mathbf{x})$ from O to G, where G denotes the center of mass of $P_0$ and $P_1$, i.e.

$$\mathbf{R}' = \mathbf{R} - \mathbf{R}_G \quad \text{or} \quad \mathbf{R}' = \mathsf{T}(\mathbf{R}_G)\mathbf{R} \tag{2}$$

where $\mathbf{R}_G = (m_0 \mathbf{R}_0 + m_1 \mathbf{R}_1)/(m_0 + m_1)$. The star position $P_0$ and the axis $GZ'$ define a plane, $\pi$, which also contains the planet $P_1$ and forms an angle $\theta$ with $GX'$ or $OX$ axis. Now, we consider the orthogonal frame of reference $Gxyz$, where $Gz \equiv GZ'$ and, subsequently, $Gx \in \pi$ and $Gy \perp \pi$. Also the positive direction of the axis $Gx$ is defined from $P_0$ to $P_1$. Thus, the reference frame $Gxyz$ is defined by a rotation, $\theta$, about the axis $GZ'$ and a position vector $\mathbf{R}' = (X', Y', Z')$ is written in the rotating frame, as $\mathbf{r} = (x, y, z)$ defined by

$$\mathbf{r} = R(\theta)\mathbf{R}', \qquad R(\theta) = \begin{bmatrix} \cos\theta & \sin\theta & 0 \\ -\sin\theta & \cos\theta & 0 \\ 0 & 0 & 1 \end{bmatrix}. \tag{3}$$

Therefore the inertial positions $\mathbf{R}_i = (X_i, Y_i, Z_i)$ and velocities $\dot{\mathbf{R}}_i = (\dot{X}_i, \dot{Y}_i, \dot{Z}_i)$ of the bodies $i$=0,1,2, are transformed for the rotating frame by the relations

$$\mathbf{r} = R(\theta)\mathsf{T}(\mathbf{R}_G)\mathbf{R}, \qquad \dot{\mathbf{r}} = R(\theta)\mathsf{T}(\dot{\mathbf{R}}_G)\dot{\mathbf{R}} + \dot{\theta} R'(\theta)\mathsf{T}(\mathbf{R}_G)\mathbf{R} \tag{4}$$

where $R'(\theta) = \partial R/\partial \theta$ and

$$\theta = \arctan\left(\frac{Y_2 - Y_1}{X_2 - X_1}\right),$$

The relations that declare the inverse of transformation (4) are written as

$$\mathbf{R} = \mathsf{T}(-\mathbf{R}_G)R(-\theta)\mathbf{r}, \qquad \dot{\mathbf{R}} = \mathsf{T}(-\dot{\mathbf{R}}_G)R(-\theta)\dot{\mathbf{r}} + \dot{\theta}\mathsf{T}(-\dot{\mathbf{R}}_G)R'^T(\theta)\mathbf{r} \tag{5}$$

According to the definition of $Gxyz$ the position of the three bodies in the rotating frame will be

$$P_1 = (x_1, 0, z_1), \quad P_2 = (x_2, y_2, z_2), \quad P_0 = (-ax_1, 0, -az_1) \qquad (a = m_1/m_0)$$

and the system in the rotating frame is described by five degrees of freedom. By taking into account the reference frame, where $\mathbf{L} \parallel Oz$, and using (5), the components of the angular momentum integral are written as

$$\begin{aligned} L_X &= \sum_{i=0}^{2} m_i (Y_i \dot{Z}_i - \dot{Y}_i Z_i) = \mu\left(b(y_2 \dot{z}_2 - \dot{y}_2 z_2) - \dot{\theta}(ax_1 z_1 + bx_2 z_2)\right) = 0, \\ L_Y &= \sum_{i=0}^{2} m_i (Z_i \dot{X}_i - X_i \dot{Z}_i) = \mu\left(b(\dot{x}_2 z_2 - x_2 \dot{z}_2) + a(\dot{x}_1 z_1 - x_1 \dot{z}_1) - b\dot{\theta} y_2 z_2\right) = 0 \end{aligned} \tag{6}$$

and

$$L_Z = \sum_{i=0}^{2} m_i (X_i \dot{Y}_i - \dot{X}_i Y_i) = \mu\left(b(x_2 \dot{y}_2 - \dot{x}_2 y_2) + \dot{\theta}\left(ax_1^2 + b(x_2^2 + y_2^2)\right)\right) = \text{const} \tag{7}$$

where

$$a = m_1/m_0, \qquad b = m_2/m, \qquad \mu = m_0 + m_1, \qquad m = m_0 + m_1 + m_2$$

The kinetic energy $T$ and the potential $V$ are written as

$$T = \frac{1}{2}\sum_{i=0}^{1} m_i \dot{\mathbf{R}}_i^2 = \frac{1}{2}\mu\left(a(\dot{x}_1^2 + \dot{z}_1^2 + x_1^2 \dot{\theta}^2) + b\left((\dot{x}_2^2 + \dot{y}_2^2 + \dot{z}_2^2) + \dot{\theta}^2(x_2^2 + y_2^2) + 2\dot{\theta}(x_2 \dot{y}_2 - \dot{x}_2 y_2)\right)\right) \tag{8}$$

$$V = -\frac{m_0 m_1}{r_{01}} - \frac{m_0 m_2}{r_{02}} - \frac{m_1 m_2}{r_{12}} \quad (9)$$

with

$$r_{01}^2 = (1+a)^2(x_1^2 + z_1^2), \quad r_{02}^2 = (ax_1 + x_2)^2 + y_2^2 + (az_1 + z_2)^2,$$
$$r_{12}^2 = (x_1 - x_2)^2 + y_2^2 + (z_1 - z_2)^2.$$

The Lagrangian of the system $\hat{L} = T - V$ does not contain $\theta$, so we get the corresponding integral $p_\theta = \frac{\partial \hat{L}}{\partial \dot{\theta}}$, which coincides with $L_z$. The integrals (8)-(9) give

$$z_1 = \frac{b}{a}\left(\frac{y_2 \dot{z}_2 - \dot{y}_2 z_2}{x_1 \dot{\theta}} - \frac{x_2 z_2}{x_1}\right), \quad \dot{z}_1 = \frac{b}{ax_1}(\dot{x}_2 z_2 - x_2 \dot{z}_2 - \dot{\theta} y_2 z_2) + \frac{\dot{x}_1}{x_1} z_1, \quad (10)$$

$$\dot{\theta} = \frac{L_z/\mu - b(x_2 \dot{y}_2 - \dot{x}_2 y_2)}{ax_1^2 + b(x_2^2 + y_2^2)}. \quad (11)$$

Therefore, we consider the Lagrange equations only for the variables $x_1, x_2, y_2$ and $z_2$ which are

$$\ddot{x}_1 = -\frac{m_0 m_2}{\mu}\left(\frac{x_1 - x_2}{r_{12}^3} + \frac{ax_1 + x_2}{r_{02}^3}\right) - \mu \frac{x_1}{r_{01}^3} + x_1 \dot{\theta}^2$$

$$\ddot{x}_2 = \frac{m}{\mu}\left(\frac{m_1(x_1 - x_2)}{r_{12}^3} - \frac{m_0(ax_1 + x_2)}{r_{02}^3}\right) + x_2 \dot{\theta}^2 + 2\dot{y}_2 \dot{\theta} + y_2 \ddot{\theta} \quad (12)$$

$$\ddot{y}_2 = -\frac{m}{\mu}\left(\frac{m_1}{r_{12}^3} + \frac{m_0}{r_{02}^3}\right) y_2 + y_2 \dot{\theta}^2 - 2\dot{x}_2 \dot{\theta} - x_2 \ddot{\theta}$$

$$\ddot{z}_2 = \frac{m}{\mu}\left(\frac{m_1(z_1 - z_2)}{r_{12}^3} - \frac{m_0(az_1 + z_2)}{r_{02}^3}\right)$$

The quantity $\ddot{\theta}$ is found by differentiating Eq. (11). Then, the term $x_2 \ddot{y}_2 - y_2 \ddot{x}_2$, which appears, is substituted from Eq. 12 (multiply Eq (12)b with $y_2$, Eq (12)c with $x_2$ and then subtract). We obtain

$$\ddot{\theta} = \frac{m_0 m_2}{\mu} \frac{y_2}{x_1}\left(r_{12}^{-3} - r_{02}^{-3}\right) - 2\dot{\theta}\frac{\dot{x}_1}{x_1} \quad (13)$$

We should note that for initial conditions $z_i(0) = 0$, $\dot{z}_i(0) = 0$ ($i$=1,2) we get planar motion with integrals the energy $E = T + V$ and the angular momentum component $L_z$.

## 2.2 Symmetric periodic orbits

We define the Poincare section plane in phase space $\hat{\pi} = \{y_2 = 0, \dot{y}_2 > 0\}$, which geometrically coincides with plane $\pi$, and choose initial conditions of orbits always on this plane.

The system (12) possesses the following four symmetries[9]

$$\Sigma_1 : (x_1, x_2, y_2, z_2, t) \to (x_1, x_2, -y_2, z_2, -t), \quad \Sigma_2 : (x_1, x_2, y_2, z_2, t) \to (x_1, x_2, -y_2, -z_2, -t)$$
$$\Sigma_3 : (x_1, x_2, y_2, z_2, t) \to (-x_1, -x_2, y_2, z_2, -t), \quad \Sigma_4 : (x_1, x_2, y_2, z_2, t) \to (-x_1, -x_2, y_2, -z_2, -t)$$

### 2.2.1. Symmetry with respect to xz plane ($\Sigma_1$).

We start an orbit from the section plane $\hat{\pi}$ and vertically to the $xz$ plane, i.e.

$$x_1(0) = x_{10}, \quad x_2(0) = x_{20}, \quad z_2(0) = z_{20}, \quad \dot{x}_1(0) = \dot{x}_2(0) = \dot{z}_2(0) = 0, \quad \dot{y}_2(0) = \dot{y}_{20} \quad (14)$$

If the above orbit crosses again the plane $\hat{\pi}$ perpendicularly at time $t = \tau$ (for first time), then the orbital arc for $0 \le t < \tau$ is mapped by the symmetry $\Sigma_1$ to the orbital arc for $\tau \le t < 2\tau$. Thus, the orbit (14) is periodic with period $T = 2\tau$, if the following periodicity conditions hold

$$\dot{x}_1(\tau; \mathbf{x}_0) = 0, \quad \dot{x}_2(\tau; \mathbf{x}_0) = 0, \quad \dot{z}_2(\tau; \mathbf{x}_0) = 0 \quad (15)$$

where $\mathbf{x}_0$ indicates the initial conditions (14). We can present a periodic orbit of $xz$-symmetry as a point in the

4D space $\Pi_1 = \{x_{10}, x_{20}, z_{20}, \dot{y}_{20}\}$.

*2.2.2. Symmetry with respect to x - axis ($\Sigma_2$).*

We start an orbit from the section *x*-axis (which belongs to the plane $\hat{\pi}$) and vertically to this, i.e.

$$x_1(0) = x_{10}, \quad x_2(0) = x_{20}, \quad z_2(0) = \dot{x}_1(0) = \dot{x}_2(0) = 0, \quad \dot{y}_2(0) = \dot{y}_{20}, \quad \dot{z}_2(0) = \dot{z}_{20} \qquad (16)$$

If the above orbit crosses again the *x*-axis perpendicularly at time $t = \tau$, similarly to the previous case, the orbit (16) is periodic with period $T = 2\tau$ and the periodicity conditions are

$$\dot{x}_1(\tau; \mathbf{x}_0) = 0, \quad \dot{x}_2(\tau; \mathbf{x}_0) = 0, \quad z_2(\tau; \mathbf{x}_0) = 0 \qquad (17)$$

where $\mathbf{x}_0$ indicates the initial conditions (16). We can present a periodic orbit of *x-axis*-symmetry as a point in the 4D-space $\Pi_2 = \{x_{10}, x_{20}, \dot{y}_{20}, \dot{z}_{20}\}$

*2.2.3. Symmetries $\Sigma_3, \Sigma_4$.*

Similarly to the previous cases we can define symmetric periodic orbits for the symmetries with respect to the *yz* plane and to the *y*-axis. According to the definition of the rotating frame, these symmetries are not defined for the *restricted* problem. In the *general planar* problem a periodic orbit with such a symmetry ought to show $x_1 > 0$ and $x_1 < 0$ (in different time intervals), thus a collision of the bodies $P_0$ and $P_1$ (*x*=0) must occur. We do not deal with such orbits in the planetary problem.

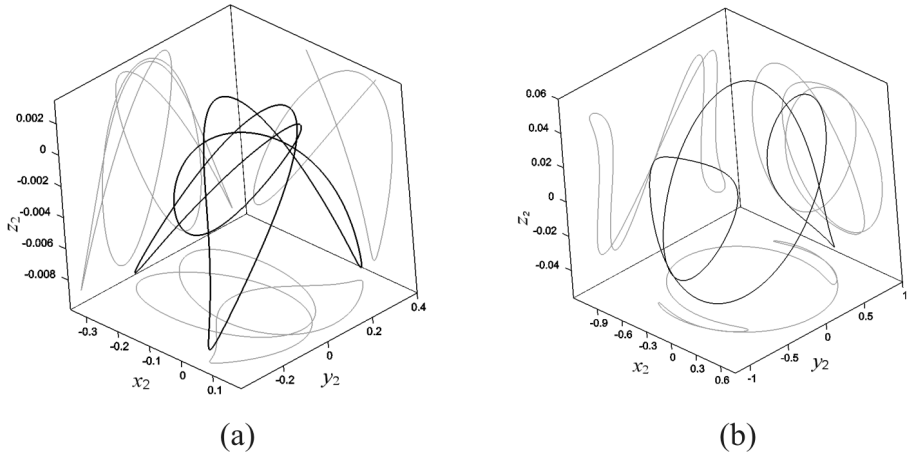

**Figure 1:** Symmetric periodic orbits in the space $x_2 y_2 z_2$ (a) *xz*-plane symmetry (b) *x*-axis symmetry, together with their projections to the planes shown with gray lines.

### 2.3 Linear Stability

We denote the 8 variables of the system $\{x_1, x_2, y_2, z_2, \dot{x}_1, \dot{x}_2, \dot{y}_2, \dot{z}_2\}$ with the vector $\mathbf{x} = (x_1, ..., x_8)$ and assume system (12) as a system of eight equations of first order with $\dot{x}_i = x_{i+4}, i = 1...4,$ and $\dot{x}_i = \ddot{x}_{i-4}, i = 5...8$. Let an orbit $\mathbf{x} = \mathbf{x}(t; \mathbf{x}_0)$ correspond to the initial conditions $\mathbf{x}_0 = \mathbf{x}(0)$. Considering a small initial deviation to the initial conditions $\mathbf{x}'_0 = \mathbf{x}_0 + \boldsymbol{\xi}_0$, $|\boldsymbol{\xi}_0| \ll 1$, we should obtain a solution $\mathbf{x}'(t) = \mathbf{x}(t) + \boldsymbol{\xi}(t)$. By substituting this solution to equations (12), after linearization, we obtain the variational equations and the corresponding solution in the form

$$\dot{\boldsymbol{\xi}} = \mathbf{J}(t)\boldsymbol{\xi} \quad \Rightarrow \quad \boldsymbol{\xi} = \boldsymbol{\Delta}(t)\boldsymbol{\xi}_0$$

where $\mathbf{J}$ is the Jacobian of the right part of system's equations and $\boldsymbol{\Delta}(t)$ the fundamental matrix of solutions which corresponds to the eight sets of initial conditions

$$\boldsymbol{\xi}_{i0} = \{\xi_{ij0}\}, \quad 1 \leq i, j \leq 8, \quad \xi_{ij0} = \begin{cases} 0 & i \neq j \\ 1 & i = j \end{cases}$$

If the solution $\mathbf{x}(t; \mathbf{x}_0)$ corresponds to a periodic solution of period *T*, then we form the monodromy matrix $\boldsymbol{\Delta}(T)$ and $\boldsymbol{\xi}(t)$ remains bounded iff all the eigenvalues of $\boldsymbol{\Delta}(T)$ lie on the unit circle. Then the orbit is called *linearly stable*. Since $\boldsymbol{\Delta}(T)$ is symplectic the eigenvalues are in conjugate pairs $(\lambda, 1/\lambda)$. Due to the energy integral, one pair of eigenvalues is always $\lambda_1 = \lambda_2 = 1$. The other three pairs can be located as it is shown in Fig. 2 and correspond to different stability types.

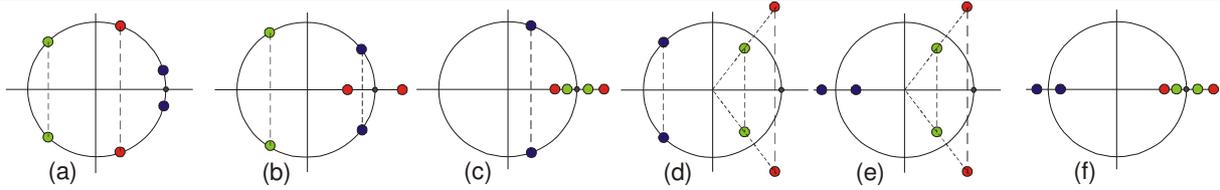

**Figure 2**: Stability types a) stability b) simple instability c) double instability d) complex instability e) u-complex instability f) triple instability.

### 2.4 Vertical Stability of planar periodic orbits

Let us consider a planar solution $\mathbf{x}(t) = \{x_1(t), x_2(t), y_2(t), \dot{x}_1(t), \dot{x}_2(t), \dot{y}_2(t)\}$ of the first 3 equations (12), where we set $z_i = \dot{z}_i = 0$, $i=1,2$. Then we isolate the last equation of Eqs. (12) for $z_2$ and we consider the initial conditions $z_{20} = \zeta_{10}$, $\dot{z}_2 = \zeta_{20}$, with $|\zeta_{i0}| \ll 1$ and the corresponding solution $\zeta_i = \zeta_i(t)$. This solution can be obtained up to first order by linearizing the last one of equations (12) with respect to the vertical components $\zeta_i$ and assuming for the planar components the solution $\mathbf{x}(t)$. Then, we obtain the following equations

$$\dot{\zeta}_1 = \zeta_2, \qquad \ddot{\zeta}_2 = A\zeta_1 + B\zeta_2 \tag{18}$$

where

$$A = -\frac{mm_0}{\mu}\left[(1-\gamma)d_{02}^{-3} + (a+\gamma)d_{12}^{-3}\right], \quad B = -\frac{m_0 m_2}{\mu}\frac{y_2}{x_1\dot{\theta}}(d_{02}^{-3} - d_{12}^{-3})$$

with

$$\gamma = b\frac{x_2\dot{\theta} + \dot{y}_2}{x_1\dot{\theta}}, \quad d_{02}^2 = (ax_1 + x_2)^2 + y_2^2, \quad d_{12}^2 = (x_1 - x_2)^2 + y_2^2$$

When the planar solution $\mathbf{x}(t)$ is $T$-periodic then (18) is a linear system with periodic coefficients and with solution of the form $\zeta = \Delta(t)\zeta_0$, where $\Delta(t)$ is the fundamental matrix of solutions corresponding to the initial conditions $\zeta_0 = (1,0)^T$ and $\zeta_0 = (0,1)^T$. The monodromy matrix $\Delta(T)$ has eigenvalues $(\lambda, 1/\lambda)$. If $\lambda$ is complex the solution $\zeta_i = \zeta_i(t)$ is bounded and the planar orbit $\mathbf{x}(t)$ is *vertically stable*. Instead, if $\lambda$ is real but not equal to $\pm 1$ the planar orbit is *vertically unstable*. The particular case $\lambda = \pm 1$ corresponds to *vertically critical orbits*.

### 2.5 Continuation of periodic orbits: from plane to space

Consider a periodic orbit of *xz*-symmetry with the initial conditions (14), which satisfy the periodic conditions (15). Let a new initial value for the variable $z_2$, $z'_{20} = z_{20} + \delta z_{20}$. Suppose that for this value we can determine new initial values $\mathbf{x}'_0 = \mathbf{x}_0 + \delta\mathbf{x}$ but the zero initial values for the variables $\dot{x}_1, \dot{x}_2, \dot{z}_2$ and the new value $z'_{20}$ are fixed. If these new initial conditions correspond to a new periodic orbit then the conditions (15) must be satisfied, namely

$$\dot{x}_1(\tau; x_{10} + \delta x_{10}, x_{20} + \delta x_{20}, \dot{y}_{20} + \delta \dot{y}_{20}) = 0,$$
$$\dot{x}_2(\tau; x_{10} + \delta x_{10}, x_{20} + \delta x_{20}, \dot{y}_{20} + \delta \dot{y}_{20}) = 0,$$
$$\dot{z}_2(\tau; x_{10} + \delta x_{10}, x_{20} + \delta x_{20}, \dot{y}_{20} + \delta \dot{y}_{20}) = 0. \tag{19}$$

The above system can be solved with respect to $\delta x_{10}, \delta x_{20}$ and $\delta \dot{y}_{20}$ by using a Newton-Raphson method. This is possible if the Jacobian of the system has nonzero determinant, i.e.

$$\begin{vmatrix} \partial \dot{x}_1/\partial x_{10} & \partial \dot{x}_1/\partial x_{20} & \partial \dot{x}_1/\partial \dot{y}_{20} \\ \partial \dot{x}_2/\partial x_{10} & \partial \dot{x}_2/\partial x_{20} & \partial \dot{x}_2/\partial \dot{y}_{20} \\ \partial \dot{z}_2/\partial x_{10} & \partial \dot{z}_2/\partial x_{20} & \partial \dot{z}_2/\partial \dot{y}_{20} \end{vmatrix} \neq 0 \tag{20}$$

The above condition generally holds and, therefore monoparametric continuation, with parameter the variable $z_2$ can be established. The generated periodic orbits form characteristic curves in the space $\Pi_1$ of initial conditions called *families*. If (20) does not hold, a bifurcation point will exist. It is proved that for $z_2=0$ we obtain a planar periodic orbit, which is vertically critical[11]. Therefore, a planar periodic orbit is continued to the third dimension, iff it is a vertically critical orbit.

In a similar way, we can continue periodic orbits with *x*-axis symmetry. In this case, the continuation parameter is the variable $\dot{z}_2$.

## 3 RESONANT FAMILIES OF PERIODIC ORBITS.

In the inertial frame the planetary orbits seem like Keplerian ellipses for short time intervals (few periods). So we can define for each periodic orbit the corresponding Keplerian orbital elements known as osculating elements. In the following, we consider the planetary eccentricities, $e_1$ and $e_2$ and present the families of planar periodic orbits on the plane of $e_1$-$e_2$. For the three dimensional orbits we set as the third dimension the mutual inclination, $\Delta i$, of planetary orbits.

### 3.1 Planar Resonant families

In Fig. 3, we present families of periodic orbits for the resonances 2:1, 5:2, 3:1 and 4:1. Although more families exist we present some of those having stable segments and can host a planetary system. Generally, the families depend on the planetary mass ratio and not on each planetary mass. So, for each resonance we present families which correspond to different mass ratios $\rho = m_2 / m_1$. In our computations, we set always $m_1$=0.001 (equal to Jupiter's mass in our normalized units, where $m$=1). The planar and vertical stability is also indicated. Blue or red line segments correspond to planar stable or unstable orbits, respectively. Solid or dashed lines indicate vertical stability or instability respectively. The vertically critical orbits are indicated by dots.

All families start from almost circular orbits ($e_i \approx 0$) and they show stable segments even for very large eccentricities. The characteristic curves are similar for the resonances 2:1, 3:1 and 4:1. Their stable segments seem to become larger as the mass ratio $\rho$ increases. Particularly, for the 2:1 resonance the families are totally stable for about $\rho > 1$. But for the 5:2 resonance the orbits for $\rho = 0.01$ are all stable. Vertical instability seems to hold for some parts of the families for moderate eccentricities. In most cases, orbits of low or high eccentricities are vertically stable.

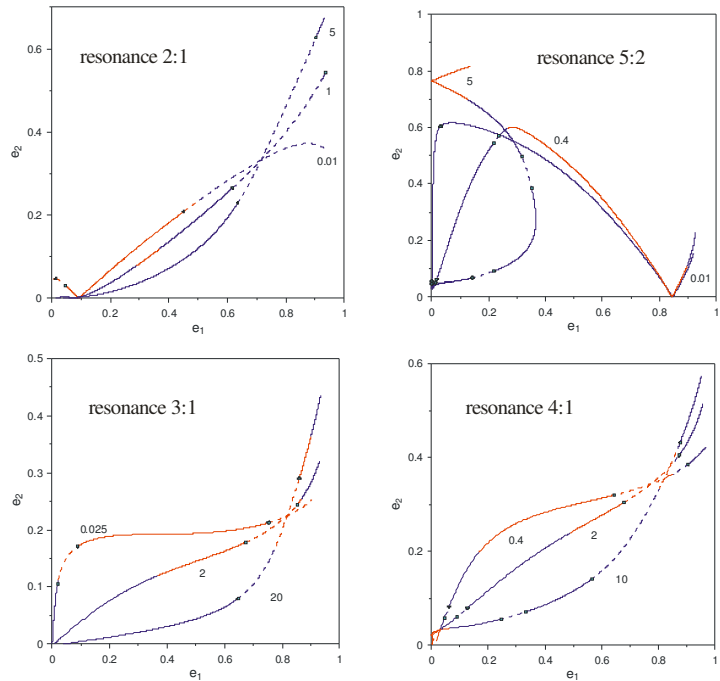

**Figure 3** Families of planetary periodic orbits in various resonances. The numbers indicate the mass ratio $\rho = m_2 / m_1$. Colors and line style indicate stability (see the text).

### 3.2 Three dimensional resonant families

Starting from the vertically critical planar orbits, shown in Fig. 3, we generate via continuation families of 3D orbits. We present these families in the space $e_1 - e_2 - \Delta i$ as it is shown in Fig. 4.

We note that stable orbits are represented by blue color and unstable ones by red color. The particular symmetry of the orbits is also indicated. A necessary condition for a 3D family to start as stable is that the planar vertically critical orbit, where it starts from, is also stable. For the resonances 2:1 and 3:1 most of the families are unstable, while the stable parts correspond to highly eccentric planetary orbits. However, in the 5:2 and, mainly, in the 4:1 resonance we can observe the existence of stable orbits for lower eccentricity values. For all resonances we can distinguish stable planetary orbits with relatively high mutual inclination (about up to 60º). However, for almost circular or low eccentric orbits stability is obtained only for $\Delta i < 15º$.

### 3.3 Planetary evolution and long-term stability

In Hamiltonian systems, stable periodic orbits are surrounded mostly by invariant tori with quasiperiodic orbits. For the planar TBP model, the numerical simulations show that in these regions long-term stability is guaranteed for the planetary orbits. The same stability seems to hold for the three dimensions. A typical example is shown in Fig. 5. Starting with initial conditions close to a stable periodic orbit we obtain that the orbital elements (particularly, we present only the eccentricities and the mutual inclination of the planets) show small amplitude regular oscillations. Instead, around unstable periodic orbits chaotic regions are formed in phase space and the evolution is very irregular (see Fig. 5). In these cases, after long-term evolution, the planets show close encounters and the system is destabilized.

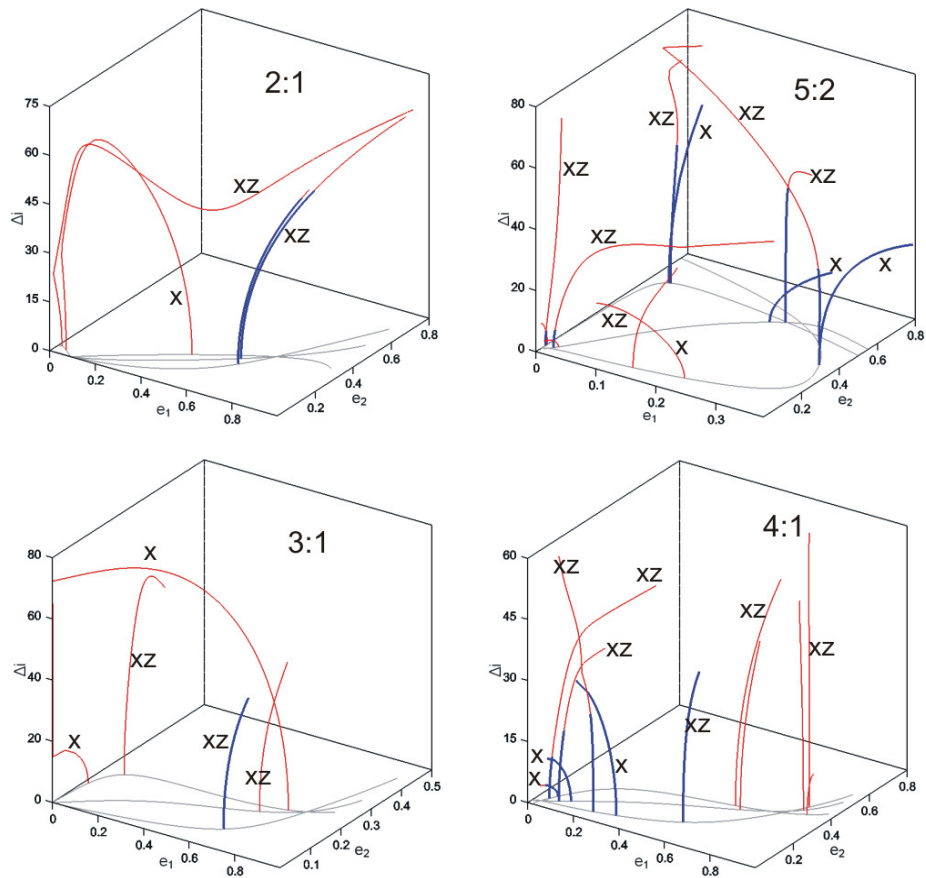

**Figure 4** Families of three-dimensional periodic orbits in the resonances 2:1, 5:2, 3:1 and 4:1. The families start from vertically critical orbits of the planar families. Blue color indicates stable orbits, while red corresponds to unstable ones.

## 4 SUMMARY.

The general three-body problem (TBP) can be used for modeling planetary systems consisting of two planets but, also can be used in understanding the dynamics of more complicated systems of many planets. In this study, we considered and presented in detail the three-dimensional case, which has been not studied sufficiently in literature. With out loss of generality, we can reduce the number of degrees of freedom of the 3D-general TBP (from six to four) by using a suitable frame of reference and by using the angular momentum integral of motion.

Periodic orbits are fundamental solutions for a dynamical system and affect the topological structure of phase space. In this work, we referred to and computed only symmetric periodic orbits, which form monoparametric families in phase space starting from vertically critical planar orbits. Their linear stability has been also computed. If a family of 3D periodic orbits starts from a planar unstable orbit then it is also unstable. However, along the family the linear stability may change. At these points bifurcations should occur to either symmetric orbits with higher multiplicity or asymmetric orbits. This is an issue for a future study.

Stable periodic orbits are very important for planetary dynamics, since they can host real planetary systems in their neighborhood. At the moment, the observations are not accurate in giving the correct positions of planets, however, many exo-solar systems seem to have been captured in resonances and since they ought to be long-term stable their position may be associated with stable periodic orbits.

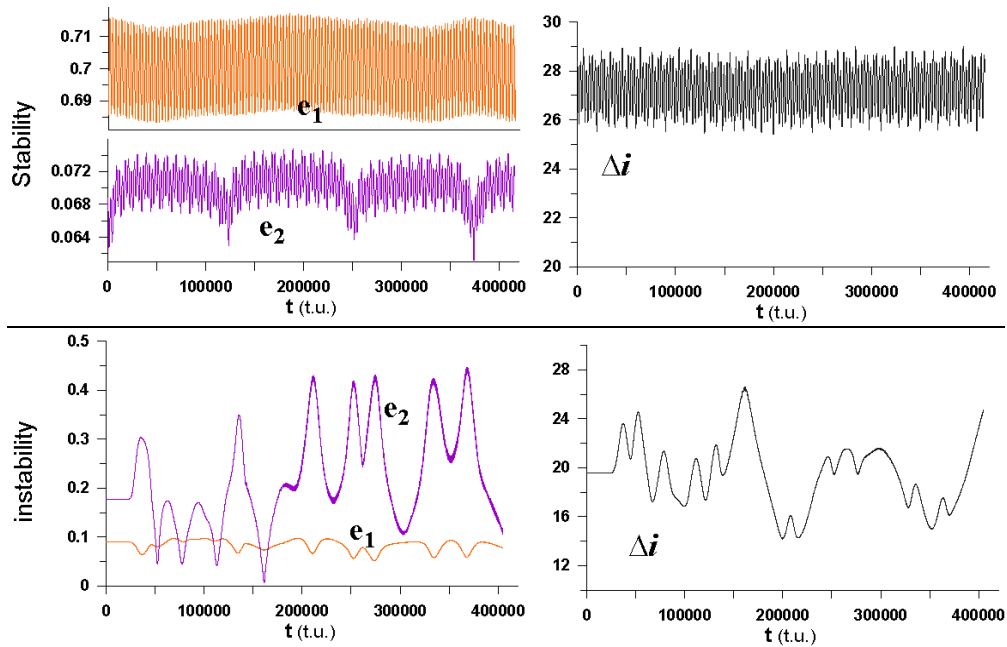

**Figure 5.** Evolution of planetary eccentricities and mutual inclination for a 3:1 resonant planetary system close to stable periodic orbit (top panels) and close to an unstable one.

**ACKNOWLEDGEMENTS.** This research has been co-financed by the European Union (European Social Fund - ESF) and Greek national funds through the Operational Program "Education and Lifelong Learning" of the National Strategic Reference Framework (NSRF) - Research Funding Program: THALES. Investing in knowledge society through the European Social Fund.